\documentclass[sn-mathphys-num]{sn-jnl}

\usepackage{graphicx}%
\usepackage{multirow}%
\usepackage{amsmath,amssymb,amsfonts}%
\usepackage{amsthm}%
\usepackage{mathrsfs}%
\usepackage[title]{appendix}%
\usepackage{xcolor}%
\usepackage{textcomp}%
\usepackage{manyfoot}%
\usepackage{booktabs}%
\usepackage{algorithm}
\usepackage{algpseudocode}
\usepackage{subfig}
\usepackage{subcaption}
\usepackage{float}
\usepackage{listings}%

\begin{document}

\title{CRUPL: A Semi-Supervised Cyber Attack Detection with Consistency Regularization and Uncertainty-aware Pseudo-Labeling in Smart Grid}


\author*[1]{\fnm{Smruti P.} \sur{Dash}}\email{202011002@iitdh.ac.in}

\author[2]{\fnm{Kedar V.} \sur{Khandeparkar}}\email{kedark@iitdh.ac.in}

\author[3]{\fnm{Nipun} \sur{Agrawal}}\email{agrawalnipun68@gmail.com}

\affil*[1]{\orgdiv{Department of Computer Science and Engineering}, \orgname{Indian Institute of Technology Dharwad}, \orgaddress{\city{Dharwad}, \postcode{580011}, \state{Karnataka}, \country{India}}}

\affil[2]{\orgdiv{Department of Computer Science and Engineering}, \orgname{Indian Institute of Technology Dharwad}, \orgaddress{\city{Dharwad}, \postcode{580011}, \state{Karnataka}, \country{India}}}

\affil[3]{\orgdiv{Department of Computer Science and Engineering}, \orgname{Rajiv Gandhi Institute of Technology}, \orgaddress{\city{Bangalore}, \postcode{560032}, \state{Karnataka}, \country{India}}}


\abstract{The modern power grids are integrated with digital technologies and automation systems. 
The inclusion of digital technologies has made the smart grids vulnerable to cyber-attacks. 
Cyberattacks on smart grids can compromise data integrity and jeopardize the reliability of the power supply. 
Traditional intrusion detection systems often need help to effectively detect novel and sophisticated attacks due to their reliance on labeled training data, which may only encompass part of the spectrum of potential threats. 
This work proposes a semi-supervised method for cyber-attack detection in smart grids by leveraging the labeled and unlabeled measurement data. 
We implement consistency regularization and pseudo-labeling to identify deviations from expected behavior and predict the attack classes. 
We use a curriculum learning approach to improve pseudo-labeling performance, capturing the model uncertainty. 
We demonstrate the efficiency of the proposed method in detecting different types of cyberattacks, minimizing the false positives by implementing them on publicly available datasets. 
The method proposes a promising solution by improving the detection accuracy to 99\% in the presence of unknown samples and significantly reducing false positives.}




\maketitle
\section{Introduction}
\label{sec1}

Integration of advanced digital technologies into the conventional power system leads to the evolution of the \textit{cyber physical system} (CPS) known as the smart grid. 
The smart grid promises more efficient energy management, improved reliability, and enhanced sustainability. 
However, the inclusion of digital technologies and increased connectivity have introduced new susceptibility in the smart grid by making it a potential target of cyber attackers.  
These cyberattacks on smart grids can have severe consequences, including service disruption, data manipulation, financial loss, and threats to public safety \cite{mitchell2014survey, yang2021survey}. 
Therefore, the cyber resilience of the smart grid is a major focus of energy providers and cyber security experts. 
Many CPSs lack security mechanisms such as message authentication, universal encryption, and dated technology, which are necessary to defend against cyberattacks such as \textit{false data injection attack} (FDIA), eavesdropping, and replay attacks \cite{dlsurvey}. 
The Stuxnet worm \cite{karnouskos2011stuxnet} and the cyberattack on the Ukrainian power grid \cite{lehman15331cyber} are examples of cyberattacks causing damage to the grid and disrupting power supply for a long duration. 

The smart grid's cyber layer comprises many sensors and \textit{phasor measurement units} (PMU) deployed to provide real-time measurements through a wide area network. 
The traditional approaches for detecting cyberattacks in smart grids often rely on signature-based \textit{intrusion detection system} (IDS). These approaches use PMU measurements to estimate the current state of the grid, where the state of a power system is the best estimate of system parameters such as voltage magnitude and angle. 
Then, the residual between the observed and estimated system parameters is computed. 
If the residual exceeds some predefined threshold, an FDI attack is declared \cite{liu2011false, kosut2011malicious}. 

In \cite{rawat2015detection}, the authors have explored attack detection based on chi-square and cosine similarity. 
The conventional model-based methods follow the rules defined according to the pattern of packets communicated in the network, the range of value of sensor readings, and some threshold defined on the system's normal behavior \cite{khan2017model}. 
These methods can effectively identify the known attack patterns or deviations in normal behavior. 
At the same time, these methods have limitations in detecting sophisticated attack patterns \cite{zhang2020detecting}. 
Moreover, the dynamic nature of smart grids poses significant challenges for traditional detection mechanisms. 
The \textit{Machine learning} (ML) techniques have emerged as promising tools in enhancing cyberattack detection in smart grids in recent years. 
There are adequate researches on cyberattack detection in smart grid using various supervised ML strategies such as \textit{support vector machines} (SVM), \textit{k-nearest neighbor} (KNN), \textit{decision tree} (DT), and \textit{random forest} (RF) that have been referred in detail in \cite{musleh2019survey, acosta2020extremely, sakhnini2019smart}. 
Moreover, supervised \textit{deep learning} (DL) algorithms such as \textit{artificial neural networks} (ANN), \textit{convolutional neural networks} (CNN), \textit{recurrent neural network} (RNN) are also employed to enhance the detection performance in \cite{niu2019dynamic, bitirgen2023hybrid}. 
The supervised algorithms rely on labeled measurement data representing known natural events and attack instances. 
However, in the real scenario, collecting labels for the attacks on smart grids is costly and time-consuming. 
Hence, to preclude the necessity of labels, the researchers investigated unsupervised ML techniques for detecting cyberattacks. 
Compared to supervised methods, the number of works using unsupervised methods is limited. The one-class SVM (OCSVM) was employed in \cite{maglaras2014ocsvm} to build an intrusion detection system (IDS) in the supervisory control and data acquisition (SCADA) system. Further, the OCSVM is combined with K-means recursive clustering to make a real-time IDS in the SCADA system \cite{maglaras2014ocsvm}. 

Due to the imbalance in the performance and availability of known instances in supervised and unsupervised methods, the semi-supervised methods became a better choice to defend CPS against attacks. Semi-supervised learning leverages labeled and unlabeled data collected from the system over time. 
The labeled data contains measurement samples during stable operations, natural events, and known attacks, while the unlabeled data includes unknown attack samples. 
The semi-supervised learning algorithms can detect cyber-attack behavior by learning from the inherent structure of data and identifying deviations from normal patterns. 
Recent research on attack detection in CPSs using semi-supervised learning focuses on building hybrid models integrating the deep learning models for data generation and a block of anomaly detection \cite{zhang2020detecting, dairi2023semi}. 
Training the deep learning models for data generation aims to generate labeled data to balance the samples of labeled and unlabeled sets. 
Thus, it helps to improve the performance of the anomaly detection module. Generating labeled data to balance the unlabeled instances adds to the time complexity of the detection method. 
Due to recent research, pseudo-labeling \cite{lee2013pseudo} has become a particularly effective technique within the realm of semi-supervised learning. 
Pseudo-labeling assigns labels to unlabeled data based on the patterns learned from labeled data by training the model iteratively. 
The iterative approach gradually improves the model's performance by integrating new information from the unlabeled data, increasing its capacity to generalize and identify unseen attack patterns.

In this work, we explore the pseudo-labeling technique to detect cyber-attacks in the grid. By taking advantage of semi-supervised approaches to learn the characteristics from a small set of labeled data, our goal is to develop an effective method for identifying both known and novel cyber-attacks. The key contributions of this work are:
\begin{itemize}
    \item Enforcing stable model prediction under input perturbation using consistency regularization.
    \item Combining consistency regularization with pseudo-labeling to determine the classes for unlabeled samples.
    \item Using curriculum learning and confidence threshold tuning to improve the model prediction and produce reliable class labels.
\end{itemize}

Consistent regularization encourages models to produce stable predictions despite input variations \cite{fan2023revisiting}. 
We iteratively train and test the model on labeled and unlabeled datasets to progressively refine the model's prediction. 
Unlike hard labels, soft labels mitigate overfitting and enhance performance under noisy conditions \cite{tang2023towards}. Hence, we consider using soft labels to retrain the model. 
Relying only on high-confidence samples can fail to capture model uncertainty. Training with high-confident samples enforces confirmation bias in the model and may lead to overfitting.  
In order to address the problem of confirmation bias, the model parameters are fine-tuned using the curriculum learning approach \cite{bengio2009curriculum}. 
Later, we calibrate the confidence thresholds of each class to determine the final pseudo-label, eliminating the risk of overfitting. 
\section{Background}
Before detailing the techniques used to construct the semi-supervised method, we review the standard operations of a CPS in this section. 
Further, we discuss the cyber-attack types targetting the CPSs and their impact. 
Moreover, the techniques used in the proposed method are introduced in this section.
\subsection{Cyber Physical System Scenario}
The typical behavior of CPSs depends on the factors such as service availability, real-time operation, and fault tolerance.  
Real-time operation is crucial for maintaining the system operation when the environment and inputs change rapidly \cite{liu2017review}. 
Finally, fault tolerance requires the system to have sufficient backups to prevent the system from shutting down during normal operation. 
The possibility of threats on the CPSs increases when they are connected to cyber-space to improve the quality of system control and service. 
The threats to a smart grid can be categorized into physical, environmental, and cyber threats. 
While the physical threats include unauthorized access to the physical equipment, environmental threats include natural disasters like extreme heat and cyclones. 
As smart grids rely heavily on the cyber layer of the system, our study is centered on cyber threats. 
Conscious attackers intentionally attack the grid to harm the grid and its operations. 
Broadly, two types of cyber attacks can compromise the security of a smart grid: passive and active attacks \cite{gunduz2018analysis}.
\subsubsection{Passive Attacks}
The hacker accesses transmitted data to learn the system's configuration and normal behavior but does not modify the content of transmitted data. Since there is no change in data, detecting these attacks is difficult. To perform passive attacks, the hacker gets unauthorized access to the system to launch various attacks, such as an eavesdropping attack where the hacker steals the data communicated between the devices without their knowledge,  
\subsubsection{Active Attacks} During an active attack, the hacker aims to affect the system's normal operation by modifying the transmitted data. To do that, the hackers perform various malicious activities, such as injecting some malware to corrupt the transmitting devices in the system, launching a denial-of-service (DoS) attack, or manipulating data known as data integrity attacks (DIA). Typically, an active attack aims to steal the data and gain control over the system.

Aligning the cyber attacks to their characteristics, we aim to detect the active attacks where the data plays a significant role in affecting the normal operations in the system. 
To ascertain the cause of disturbance in the system's normal operation, we need to investigate the change in the pattern of communicated data throughout the system. 
To accomplish this task, we need to analyze a large set of data in CPSs, such as smart grids.  
As discussed in section~\ref{sec1}, it is difficult to indicate the characteristics of current system data concerning different types of attacks. 
A large set of measurement data with unknown event instances is encountered, which is difficult to classify using the supervised ML models. 
Hence, semi-supervised methods utilize a small set of known attack instances to correctly recognize the types of attacks in the measurement data. 
The class labels of the known attack instances are exploited to define the labels for the unknown instances using pseudo-labeling (PSL). 
Some recent works on semi-supervised learning for anomaly detection focused on using pseudo-labeling to efficiently learn anomalous patterns in data and distinguish those \cite{sohn2020fixmatch, arazo2020pseudo, zhang2021flexmatch}.

\subsection{Pseudo-labeling (PSL)}
\label{psl}
\begin{figure}[!ht]
    \centering
    \includegraphics[height = 60mm, width = \textwidth]{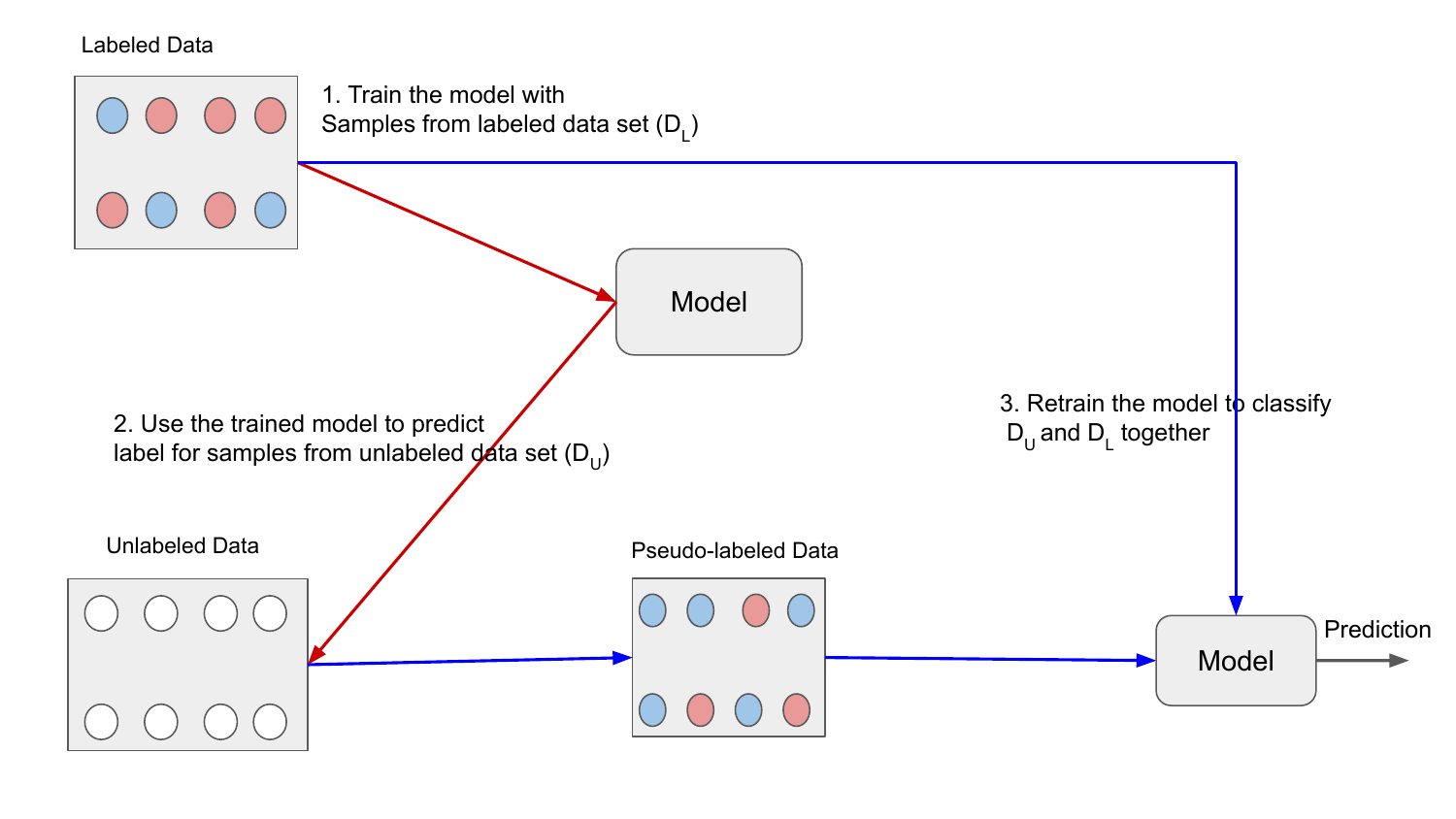}
    \caption{Pseudo-labeling}
    \label{fig:psl}
\end{figure}

Pseudo-labeling is based on a self-training framework, where a model $M$ undergoes multiple training iterations, using its prior knowledge to improve performance in subsequent steps \cite{li2019naive}.
In SSL applications, commonly the dataset consists of two distinct sets of data: labeled data $D_l(X_l, Y_l) = \{x_i^l, y_i^l| i =1,2,\cdots N_l\}$ and an unlabeled data $D_u(X_u) = \{x_i^u|i =1,2,\cdots N_u\}$. 
Here, $x_i^l \in D_u$ represents the inputs with corresponding labels $y_i^l$, while $x_u \in D_U$ represents the inputs without labels.
Typically, $|N_l|\ll |N_u|$.
Initially, the model is trained on the labeled subset $D_l$. 
Before the next iteration, the trained model is used to predict approximate labels, known as pseudo-labels $\Tilde{Y}_u$, for the unlabeled input $X_u$. In the following iterations, the model is retrained on the combined dataset $\hat{D} = \{\hat{x_i}, \hat{y}_i|i = 1,2,\cdots, (N_l+N_u)$, where $\hat{x_i}$ belongs to $(X_l \cup X_u)$ and $\hat{y}_i$ belongs to $(Y_l,\Tilde{Y}_u)$. Fig.~\ref{fig:psl} illustrates the concept of pseudo-labeling.
Pseudo-labeling provides pseudo-labels for unlabeled samples to guide the learning process. 
An early attempt of pseudo-labeling in \cite{lee2013pseudo} uses the neural network's prediction as labels for the samples. 
They pre-train the network to initialize and constrain the pseudo labels. Later in \cite{shi2018transductive}, the pseudo-labeling approach introduces an uncertainty loss for the samples with distant k-nearest neighbours in the feature space. 
Some recent works on pseudo-labeling \cite{sohn2020fixmatch} use consistency regularization with pseudo-labeling to stabilize the model's prediction in the presence of noise. 
The consistency regularization ensures that the model should output similar predictions for perturbed versions of the same sample. 
The final label of the unlabeled instances is decided based on a threshold on the confidence score of the samples.
Threshold-based pseudo-labeling guarantees the acquisition of artificial labels with the largest class probability above a predefined threshold \cite{lee2013pseudo}. 
A pseudo label $\Tilde{y}_i^u$ for a unlabeled sample $x_i$ is assigned as follows:
\begin{equation}
    \label{eq:psl}
      \Tilde{y}_i^u = \begin{cases} 
      argmax_{c\in 1,2,\cdots,C}\ p_{i}^u & \text{if $max_{c}\ (p_i^u) > \tau$}\\
      ignore & otherwise
    \end{cases}
\end{equation}
Where $\Tilde{y}_i^u$ is the class label assigned as a pseudo label to $i$th unlabeled sample, $p_{i}^u$ is the class probability vector for $i$th unlabeled sample, and $\tau$ is the threshold on confidence score to decide the class label. $C$ is the number of classes.
\subsection{Consistency Regularization}
\label{consreg}
Consistency regularization is a technique that encourages neural networks to make consistent predictions that are invariant to perturbations \cite{fan2023revisiting}. 
For a $c$-class classification problem, let say $X = \{(x_l, y_l): l \in (1,\cdots , B)\}$ is a batch of $B$ labeled examples, where $x_l$ are the training examples and $y_l$ are the one-hot labels. 
The model is trained on augmented data to ensure consistency of model prediction. 
The works in literature perform two types of augmentations, such as weak and strong augmentation, on data to train the model \cite{sohn2020fixmatch, fan2023revisiting}. 
Weakly augmented data is formed by adding relatively less noise to the training data than the strongly augmented data. 
The model is trained on the batch of training data via an unsupervised loss to generate augmented data.
\begin{equation}
    \label{eq:conreg}
    \frac{1}{B}\sum_{i=1}^{B} \lVert p(y|(x_i)) - p(y|\alpha(x_i))\rVert_2^2
\end{equation}
where $\alpha$ is the augmenting function. $p(y|x_i)$ is the class distribution predicted by the model for input $x_i$.
\section{Proposed Method: Consistency Regularization and Uncertainty-Aware Pseudo-Labeling (CRUPL)}
The CRUPL is a combination of consistency regularization and pseudo-labeling. 
We use the consistency regularization described in section~\ref{consreg} and the pseudo-labeling reviewed in section~\ref{psl}. 
The model is trained with the original and augmented labeled data via a supervised loss $l_s$. 
Typically, $l_s$ is a standard cross-entropy loss given as
\begin{equation}
\label{eq:cross}
    l_s = -\sum_{i=1}^c y_i \log(p(y_i|x))
\end{equation}
where $p(y_i|x)$ is the softmax probability predicted by the model. 

The unlabeled measurement data that we are to classify are the time series signals captured by the sensors in the grid.
Again, the labeled data, which helps to determine the class labels for the unlabeled data, are also time series data.
Hence, we formulate the neural network model for pseudo-labeling as a temporal CNN model (TempCNN)\cite{pelletier2019temporal}. 
The temporal convolutional neural networks (TempCNN) are efficient in time series processing in comparison to the RNNs \cite{lin2020temporal}.
The model parameters, such as weight and biases, are optimized using the standard cross-entropy loss as in Eq-~\ref{eq:cross}. 
The model comprises three temporal convolutional (TCN) blocks, a dense and a softmax output layer. 
The softmax layer predicts the softmax probability $p(y_i|x)$ of each sample $x$ to be in each class $y_i$.
\subsection{Temporal Convolutional Neural Networks:}
A CNN is usually composed of two parts. In part 1, convolutions and pooling layers are used alternatively to generate deep features from input. 
In part 2, a \textit{multilayer perceptron} (MLP) network is connected to classify the generated features. 
Each layer has a specific structure: (i) The input layer receives the time series of length $N$ as input. 
It has $N\times k$ neurons, where $k$ denotes the number of input time series. 
(ii) A convolutional layer performs convolution operation on the time series of the preceding layer with convolution filters. 
The selection of parameters, such as the filter size, the number of filters, and convolution strides, is based on the experiment. 
(iii) The pooling layer performs an operation that downsamples the convolutional output. 
After several convolution and pooling operations, the original time series is represented as a series of feature maps. 
All the feature maps are connected in the Flatten layer. 
(iv) The output layer has $n$ output neurons corresponding to $n$ classes. 
It is fully connected to the flatten layer. 
The output of this layer is the vector of softmax probabilities of the preceding layer output $z$ concerning each class computed using Eq. ~\ref{eq:softmax}.
\begin{equation}
    \label{eq:softmax}
    softmax(z)_i = \frac{e^{z_i}}{\sum_{j+1}^N e^{z_j}}
\end{equation}

\begin{figure}[!ht]
    \centering
    \includegraphics[height = 50mm, width = \textwidth]{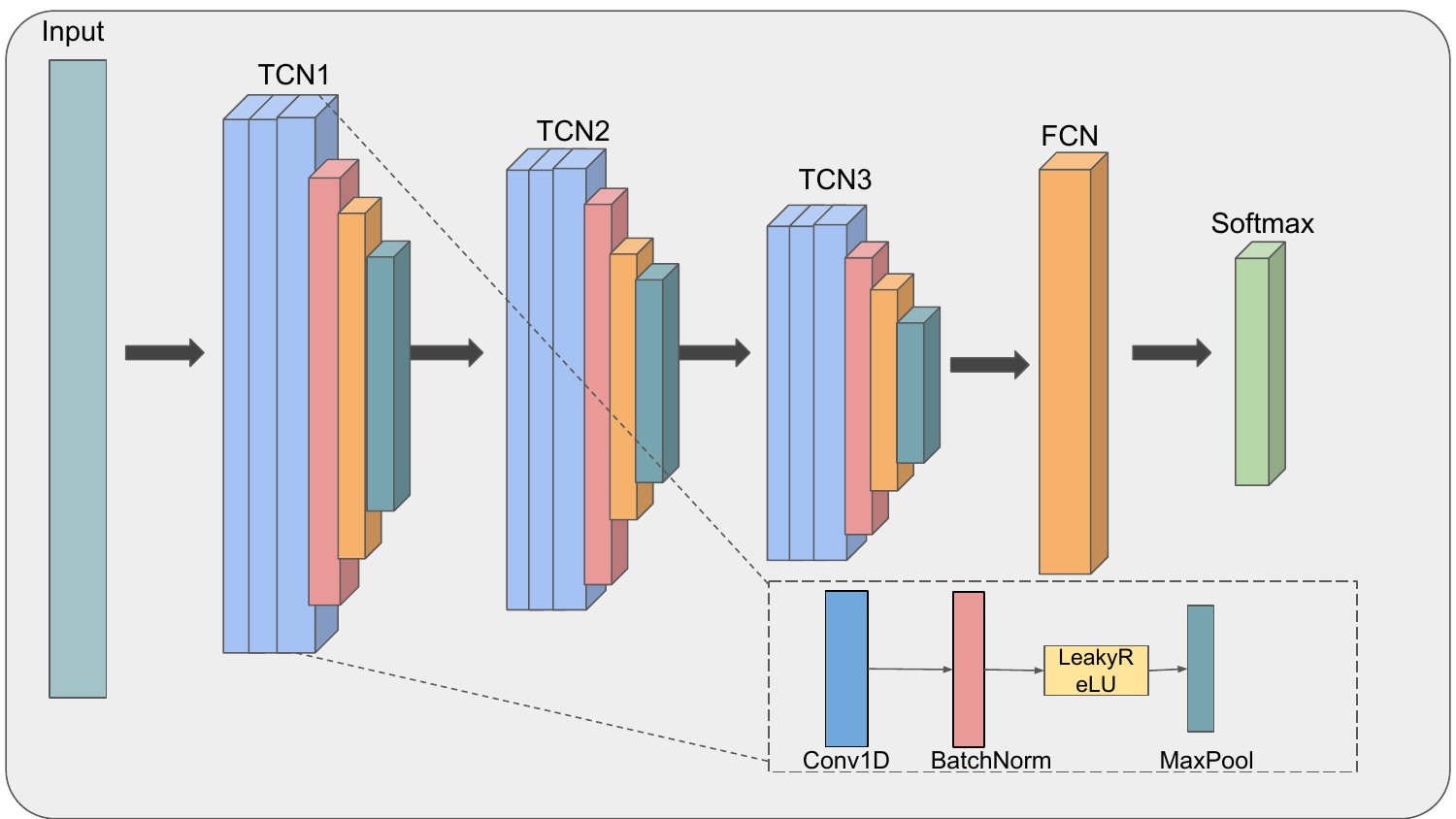}
    \caption{The Temporal CNN model used for Pseudo-labeling}
    \label{tempcnn}
\end{figure}

In order to process the time series signal, we use the 1-dimensional CNNs (1D-CNN) \cite{kiranyaz20211d} called temporal CNNs. 
The 1D-CNNs are well-suited for real-time and low-cost applications because of their low computational requirements. 
These networks typically consist of temporal convolutional (TCN) blocks, combining convolution, pooling, and batch normalization layers. 
The architecture of the TempCNN used in the proposed method is illustrated in Fig.~\ref{tempcnn}

The proposed method is illustrated as algorithm~\ref{crupl}.
In the first step of the algorithm, we train the TempCNN model denoted as $M$ on the limited labeled data.
The weakly augmented data is generated by adding noise to the labeled data.
To ensure prediction consistency, the model is trained and tested on the augmented labeled data considering the class labels of originally labeled data.

In the next step for fine-tuning the parameters, (i) the trained model predicts the softmax probability for each sample being in each class.
These probabilities are considered soft labels and used for further training.
The training using the soft labels outperforms the use of hard labels \cite{arazo2020pseudo}. 
(ii)Instead of computing the hard labels based on some threshold, we use the soft labels generated by the softmax model to train the model in the following iterations.  
In this step, we train the model iteratively to improve the prediction. 
We progressively augment the training data throughout the iterations.
In the first iteration, we include the high-confidence samples in the training data for the respective class and train the model.
In the successive iterations, we add the low-confident samples to the training data and train the model.
(iii) The freshly trained model predicts the soft labels for unlabeled samples in the next iteration. 
The loop at step $5$ is repeated for a constant $n$ number of times. 

\begin{figure}[ht]
    \centering
    \includegraphics[height = 50mm, width= \linewidth]{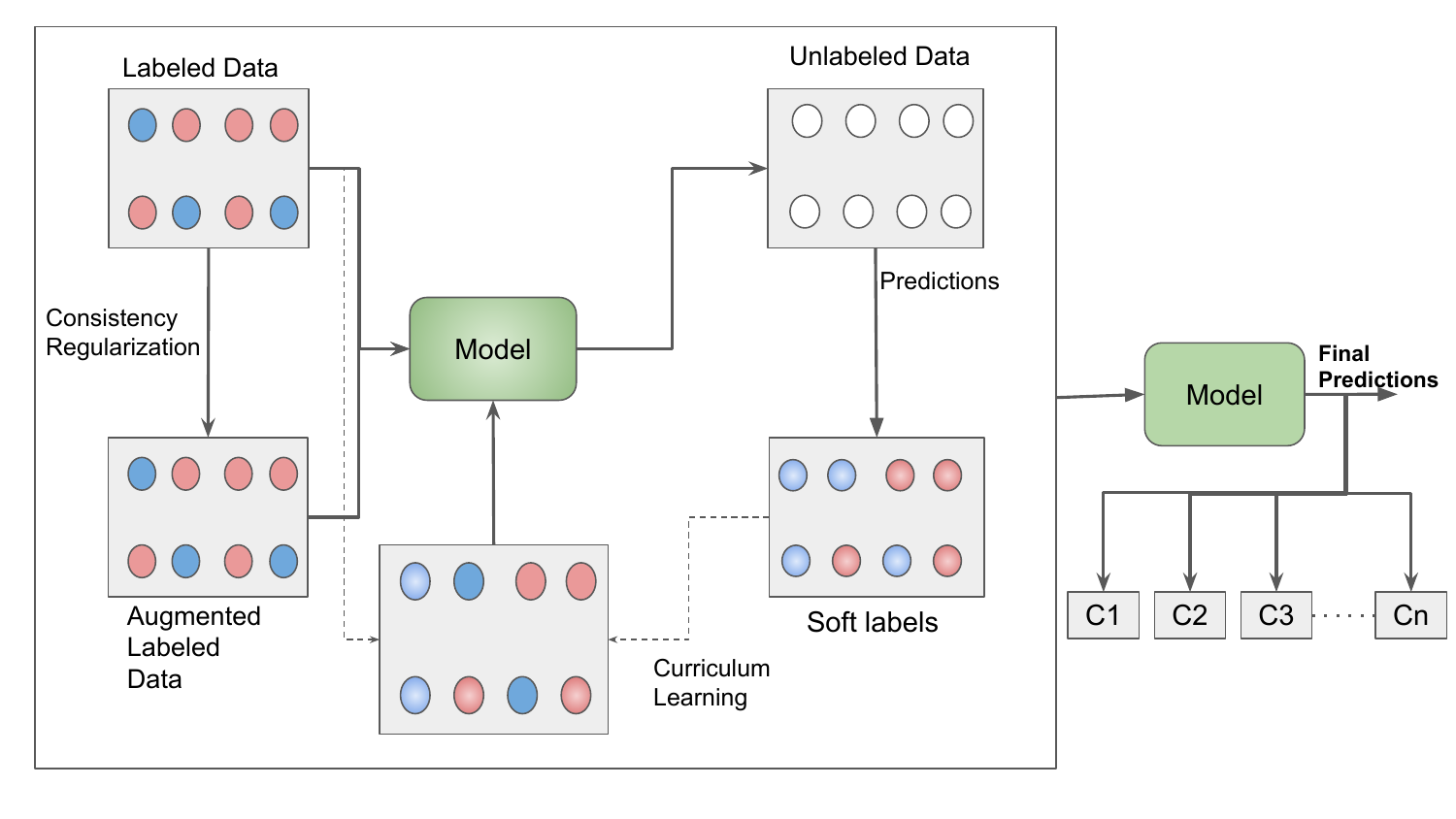}
    \caption{Pseudo-Labeling with Consistency Regularization and Curriculum Learning}
    \label{fig:crupl}
\end{figure}

Finally, the fine-tuned model predicts the class probabilities for each unlabeled sample in step $6$. 
In the next step (pseudo-labeling loop), the final class label for each sample is decided based on a threshold on the class probabilities. 
For this purpose, we use dynamic thresholding by selecting the 90th percentile value as the threshold for each class.
Later, we tune each class's confidence threshold based on each class's evaluation accuracy.
Classes with lower validation accuracy are assigned lower thresholds, encouraging the model to compute pseudo-labels closer to the model's predicted outputs.
\begin{equation}
    \label{threshold}
    \mathbf{\tau}(c) = a(c).\tau_c
\end{equation}
Here, $\mathbf{\tau}(c)$ represents the updated threshold for class $c$, while $a(c)$ denotes the evaluation accuracy of class $c$. 
Moreover, $\tau_c$ is the $90$th percentile value of the probability vector corresponding to that class. The steps of the algorithm~\ref{crupl} are illustrated in Fig.~\ref{fig:crupl}

\begin{algorithm}[H]
    \caption{Algorithm to Implement CRUPL}
    \label{crupl}
    \begin{algorithmic}[1]
        \Require $\{(D_L:(X_L,Y_L), D_{U}:(X_U))\}$
        \State  $M.fit(X_L, Y_L)$
        \State  \text{Generate Weakly augmented Data} $D'_L: (X'_L,Y'_L)$
        \State  $M.fit({X'_L},{Y'_L})$
        \State  $t = 75$
        \State $x_{train} = X_L$
        \State $y_{train} = Y_L$
        \While{$t<100$}  (\emph{Loop for Fine-tuning})
            \State $yhat = M.predict(X_{U})$
            \State $CI = Confidence\_Interval(yhat,t)$
            \For{$x \in X_{U}$} (\emph{Updating Training Data})
                \If{$yhat[x] \in CI$}
                    \State $x_{train} = x_{train}\cup x$
                    \State $y_{train} = y_{train}\cup yhat[x]$
                \EndIf
            \EndFor
            \State $M.fit(x_{train}, y_{train})$
            \State $t = t+5$
        \EndWhile
        \State $yhat = M.predict(X_{U})$
        \For{$x \in X_U$} (\emph{Pseudo-Labeling loop})
            \State $m = max(yhat[x])$
            \If{$m>\tau(c)$}
                \State \text{Assign label c for} $x$
            \EndIf
        \EndFor
        \State \text{Return labels for all samples in} $X_{u}$
    \end{algorithmic}
\end{algorithm}

The algorithm builds a reliable model that yields trustworthy class labels for the unlabeled samples. 
Moreover, the asymptotic time complexity of the algorithm becomes linear in terms of the size of unlabeled data $O(N_u)$. 
The trained model predicts the class label in constant time $O(1)$ for a new incoming sample.
\section{Experiment Analysis}
This section discusses the datasets and presents a detailed analysis of the experimental results. 
First, we discuss the structure of datasets, including
the types of events and attacks simulated, features considered, and size of datasets.  
Then, we present the experimental results, examining key metrics, trends, and observed patterns. 
Further, we illustrate the findings, their implications, and potential applications towards the research objective.
\subsection{Datasets}
This section is devoted to a detailed description of datasets adopted to assess the efficiency of the proposed method. 
We use the publicly available datasets \cite{1trw-n685-22} and \cite{morris2014industrial} to implement the proposed method and evaluate its performance.
\subsubsection{Dataset 1}
\label{data1}
The dataset has several smart grid communication data files. 
It has three different datasets, BUT-IEC104-I, VRT-IEC 104, and GICS-MMS, containing normal traffic data and different types of attacks. 
The dataset BUT-IEC104-I contains the traffic data from a real smart grid supporting the industrial network standards IEC 608705-104 and was generated from Brno University of Technology. 
The dataset VRT-IEC 104 contains traffic data due to different attacks on IEC 104 communication created using an IEC virtual testbed developed in the same University. 
Further, the dataset GICS-MMS contains the traffic data of attacks on MMS communication created manually by the G-ICS labs at the University of Alpes, France.

More details on the dataset can be found in \cite{1trw-n685-22}. 
We implement the proposed method on the first dataset, BUT-IEC104-I. This dataset was created by Matousek et al. \cite{matouvsek2020flow}.  
The dataset includes several features such as IP addresses, ports, object ID, and other derived features such as start time, end time, and quantity of exchanged data \cite{1trw-n685-22}. 
Besides the traffic under normal state, the following are the attack scenarios that were created to gather the traffic during attacks:
\begin{itemize}
    \item \textbf{Connection loss attacks:} In this case, a connection failure was implemented, due to which a short blackout happened in a device. In the first connection failure for $10$ minutes, $146$ packets were lost. The second connection loss for one hour causes losses of $921$ packets. 
    \item \textbf{DOS attack against IEC 104 control station:} The goal of a DOS attack here is to crash a control station and collapse the grid. The attacker uses a spoofed IP address to get access to the victim and floods it with $1049$ messages in $30$ minutes.
    \item \textbf{Injection commands attacks:} 
    In this case, the attacker sent unusual requests by compromising a host in the ICS network. 
    In the first phase, the attacker sent $83$ activation messages to execute the intended false commands on the target host for 5 minutes. 
    In the next phase, the attacker tried to send a file from the target host to the compromised one. 
    \item \textbf{Rogue devices attack:} 
    This attack occurs when some unauthorized devices are connected to the network and communicate with hosts using legitimate messages. 
    Here, in this scenario, a rogue device was connected to the ICS network and sent spontaneous messages to a host to which the host replied with supervisory signals. 
    This attack duration was $30$ minutes and $417$ packets were sent in that duration. 
    \item \textbf{Scanning Attack:}
    In this scenario, the attacker performs a horizontal scanning of the IPs in the network. 
    Then sent IEC 104 Test Frame messages on port 2404. 
    Upon getting responses from the network stations it performs a vertical scan of the host using General Interrogation ASDUs \cite{matouvsek2021efficient}. 
    This attack lasted about 15 to 20 minutes.
    \item \textbf{Switching attack:} A malware-based attack intends to switch on/off the target device. In this attack, a series of $72$ IEC 104 packets were sent to the target in an interval of $10$ minutes, causing the device to turn on/off.
\end{itemize}
\subsubsection{Dataset 2}
\label{data2}
\begin{figure}[ht]
    \centering
    \includegraphics[height = 60mm, width = \textwidth]{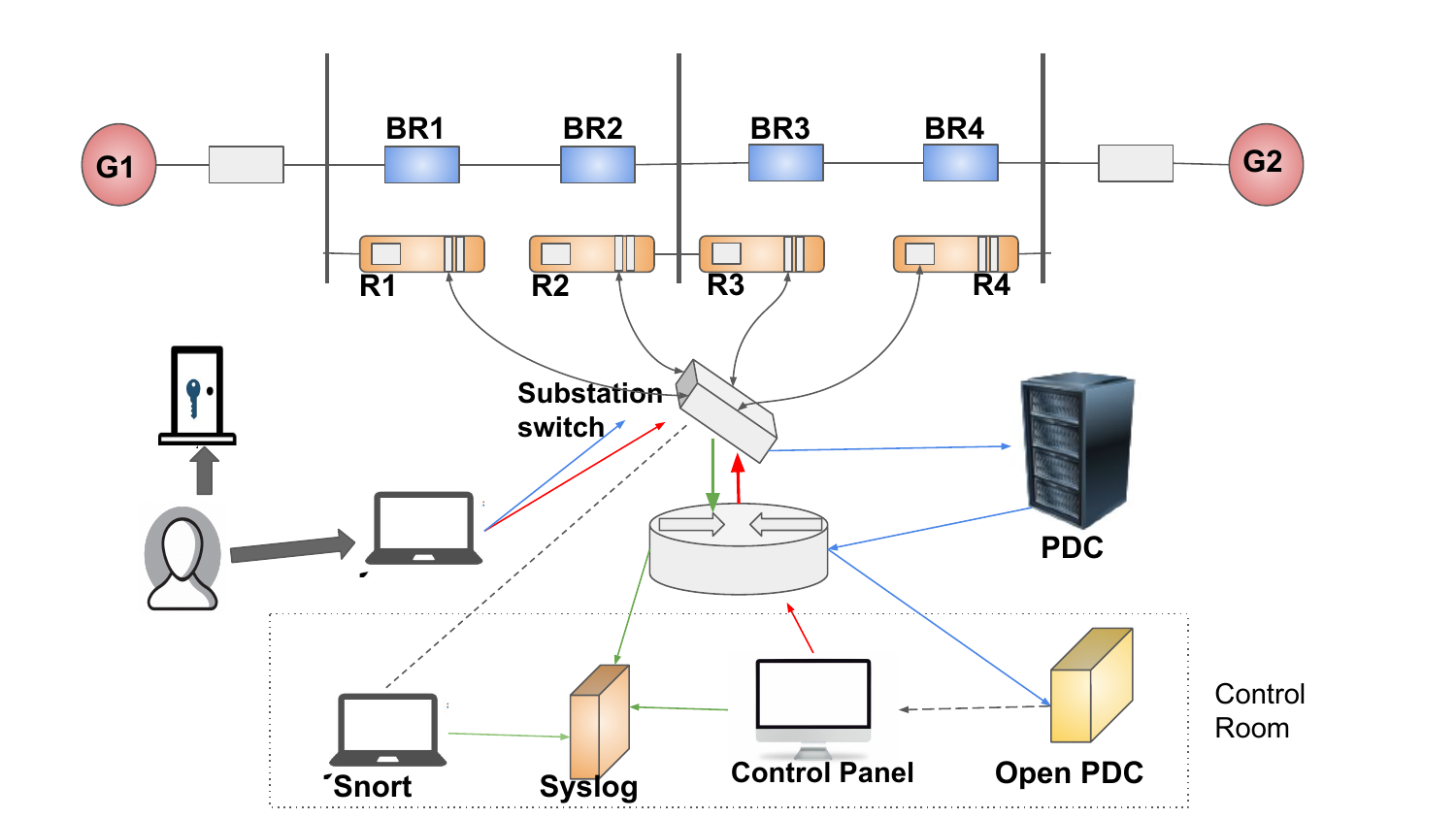}
    \caption{Power System Framework \cite{hink2014machine}}
    \label{fig:ps}
\end{figure}

This dataset is generated by Tommy Morris et al. \cite{tommymorris}. Three datasets are made from one initial dataset of fifteen sets with $37$ power system event scenarios each. 
In the $37$ scenarios, one scenario is for the stable state of the system without any event occurring, $8$ scenarios are created for $8$ natural events, and $28$ scenarios of attacks are created. 
The power system on which the scenarios are created is illustrated in Fig.~\ref{fig:ps}. 
The network has several interconnected components, such as generators, \textit{intelligent electronic devices} (IEDs), circuit breakers, and transmission lines. 
The power generators $G1$ and $G2$ are the power providers. 
The IEDs $R1 - R4$ are toggled to switch the breakers $BR1 - BR4$ on and off respectively. 
These IEDs use a distance protection scheme to toggle the breakers when a fault occurs at some point in the system. 
Whereas the IEDs have no internal validation, the breakers toggle, regardless of whether the fault is natural or attacker-created. 
There are two transmission lines, $L1$ and $L2$, that connect the breakers $BR1 - BR2$ and $BR3 - BR4$, respectively. 
The measurement signals are collected under different operational conditions for broadly three categories of events: no events, natural events, and attack events. 
In the category of normal events, the short-circuit fault and line maintenance are simulated. 
For the attack category, three types of attacks are simulated under different operational conditions at different points of the system. 
The three types of attacks include remote tripping command in injection, relay setting change, and data injection attack. 
There are $4$ PMUs deployed in the system that are integrated with the IEDs. 
$29$ features are collected from each PMU; hence, in total, $116$ features are collected.  
As the focus of our study is to use PMU data to detect cyber attacks, additional cyber-domain features collected by the system from the log information of the control room were not included. 
The attack and event scenarios created for this dataset are explained as follows:
\begin{itemize}
    \item \textbf{Short-circuit fault:} A short circuit fault occurs if there is an abnormal connection between two points with different voltages. 
    It can occur in various locations along a transmission line. 
    The event is simulated on different lines with different percentage ranges that indicate the location.
    \item \textbf{Line maintenance:} To simulate this event one or more relays are disabled for each line maintenance. 
    \item \textbf{Remote Tripping Command Injection Attack:} The attacker sends false fault commands to relays on different lines at different locations to open the breaker unexpectedly. 
    \item \textbf{Relay Setting Change Attack:} Here, the attacker changes the distance setting of the relays to disable the relay function. As a result, the relays will not trip for a valid fault or command. 
    \item \textbf{Data Injection Attack:} The attacker imitates a valid fault by modifying the communicated data such as current, voltage, and sequence components. 
    This attack intends to fool the operator and cause a blackout.
\end{itemize}
\subsection{Experimental Setup}
As per the implementation requirement, we need a small set of labeled data and a large set of unlabeled data. 
Hence, we split the datasets into two subsets, of which $5\%$ of the entire dataset is considered as labeled data, and the class labels of those are retained. We ignore the labels for the rest of the $90\%$ of the entire dataset and consider those as unlabeled data. We prepare $6$ different data sets, each size $58700$, from the Dataset $1$, each containing a different percentage of attack and normal traffic data. 
These sets are designed to examine the method's performance with varied percentages of attack data in the sensor reading. Further, 15 sets of data from Dataset 2 are collected from the public repository of the dataset provided by \cite{morris2014industrial}. The steps followed for training the model with the datasets are as follows:
\subsection{Evaluation Metrics}
\label{metrics}
The feature values of the training dataset are used to train the ML models. 
The feature values of the testing dataset are used to evaluate the models' performance through measures such as \textit{accuracy, precision, recall, F1-score}. 
To define these metrics, we take the help of true positives (TP), which have original and predicted labels as positive; true negatives (TN), which have both original and predicted labels as negative; false positives (FP), which have an original label that is not positive but the predicted label is positive, false negatives (FN) which have an original label that is not negative but the predicted label is negative. 
\begin{itemize}
    \item The precision explains how many correctly classified samples turned out to be positive and can be expressed as \cite{hossin2015review}
        \begin{equation}
            Precision=\frac{TP}{TP+FP}
        \end{equation}
    \item The recall explains how many actual positive cases the model was able to predict correctly and can be expressed as \cite{hossin2015review}
        \begin{equation}
            Recall=\frac{TP}{TP+FN}
        \end{equation}
    \item The F1-Score is the harmonic mean of precision and recall. it is maximum when both are equal and can be expressed as \cite{hossin2015review}
        \begin{equation}
            F1-Score=2\times\frac{Precision\times Recall}{Precision+Recall}
        \end{equation}
    \item The Accuracy measures how often the classifier correctly predicts. We can define accuracy as the ratio of the number of correct predictions and the total number of predictions as \cite{hossin2015review}
    \begin{equation}
            Accuracy = \frac{TP+TN}{TP+TN+FP+FN}
    \end{equation}
    \item The false positive rate (FPR) measures positive cases incorrectly classified as positive. It is the probability that a false alarm will be raised and can be defined as \cite{tharwat2021classification}
    \begin{equation}
        FPR = \frac{FP}{FP+TN}
    \end{equation}
\end{itemize}

The TempCNN classifier is trained on the $5\%$ labeled and weakly augmented labeled data prepared by adding noise to the labeled data. Then, the trained classifier is used to predict softmax probabilities for the unlabeled data. 
In each iteration, the model predicts soft labels for each sample, representing the degree of membership.  
To account for prediction uncertainty, we identify samples whose degree of membership falls within the confidence interval defined in the corresponding iteration and add them to the training data as instances of their respective classes. 
The model is then retrained with the updated training set, and soft labels are again predicted for the unlabeled data in the next iteration. 
Upon reaching the maximum number of iterations, we finalize the soft labels as the final degree of membership and compute the hard labels. 
The maximum number of iterations is decided empirically to be $10$. 
Additionally, defining a flexible threshold for each class ensures that generated pseudo-labels are reliable. 
Consequently, the labels, denoted as $\Tilde{y}_i^u$, are defined as follows:
\begin{equation}
    \Tilde{y}_i^u = argmax(p_i^u \geq \tau_c)
\end{equation}
where $p_i^u$ is the probability vector of $x_i$ and the threshold for each class $c$ is $\tau_c$.
\subsection{Results}
\label{sec:res}
\begin{figure*}[h]
    \subfloat[]{\includegraphics[height = 60mm, width =0.5\linewidth]{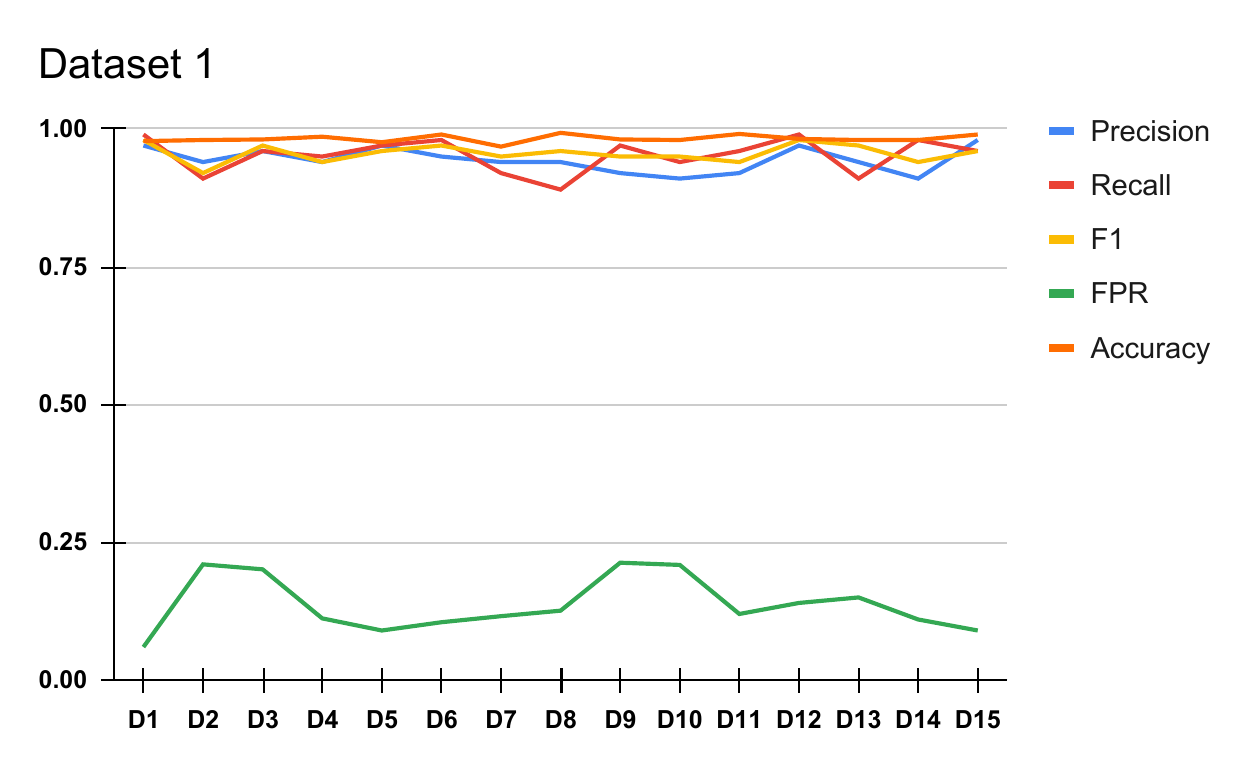}%
    }
    \subfloat[]{\includegraphics[height = 60mm, width=0.5\linewidth]{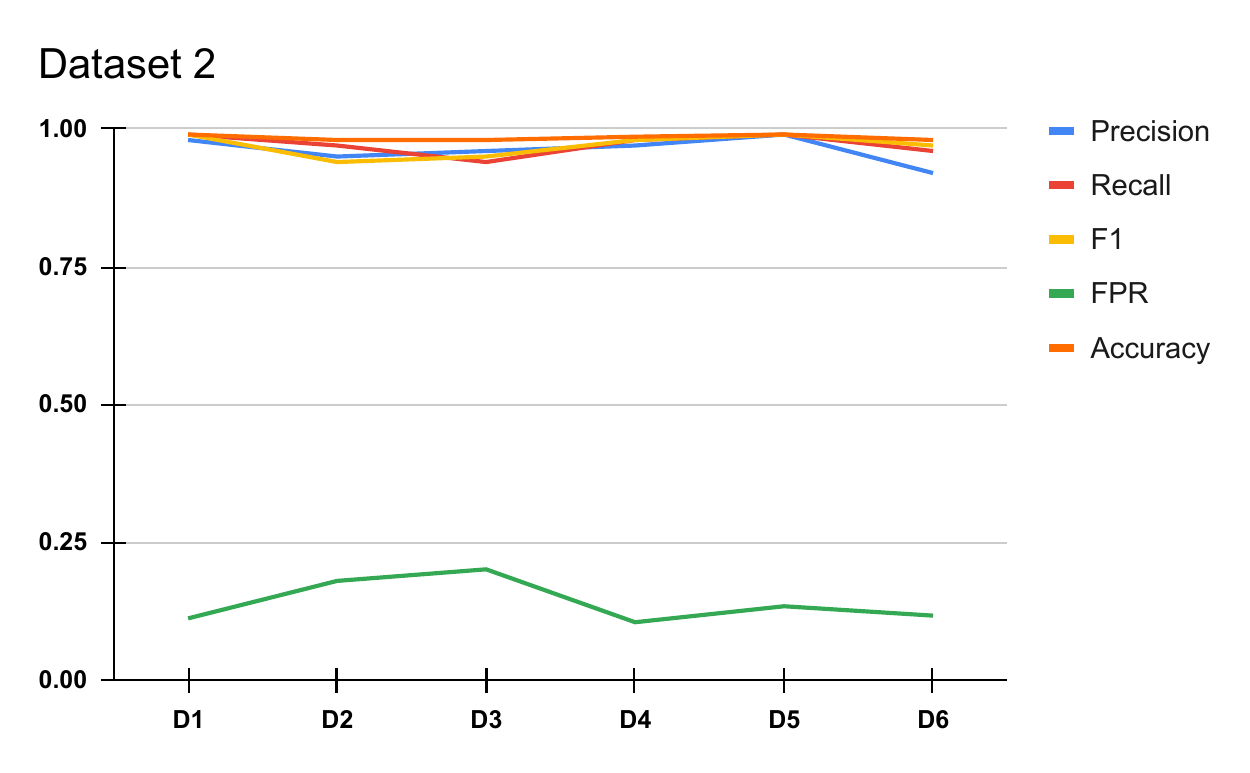}%
    }
    \caption{Performance of Proposed Method on (a) Dataset 1 and (b) Dataset 2}
    \label{results}
\end{figure*}

Here, we present the results of the proposed pseudo-labeling method for generating labels for unlabeled samples. 
Following the training on both datasets, the model's performance was assessed using various metrics, including detection accuracy, false positive rate (FPR), and additional measures detailed in Section~\ref{metrics}. 
Fig.~\ref{results} illustrate the experimental results on the sets of both datasets.

The proposed method achieves a high detection accuracy across both datasets, demonstrating the efficacy of the method in identifying cyber-attacks. 
The inclusion of a sample based on increasing confidence interval for training the model reduces the confirmation bias in the model. 
With the steps followed in the method, the model can adapt the complex patterns in smart grid data, reaching an accuracy of at least $98\%$ on both datasets.  
This result highlights the advantage of the proposed method, which reduces the confidence bias and improves the model's generalization to unseen samples.

By setting a flexible threshold for each class, the model selectively incorporated the class-wise learning effect, reducing the misclassification of benign data as attacks. 
This thresholding technique significantly dropped the FPR, making the model more reliable.

The model's performance on unknown samples further illustrates its robustness. 
The proposed method allowed the model to generalize beyond labeled instances, effectively identifying novel cyber-attack patterns not present in the initial labeled data. 
The model maintained its high accuracy even when tested against unknown measurement data, underscoring the adaptability of the proposed method.

\subsection{Summary of Key Metrics}
Table \ref{tab:tommoris} provides a summary of key metrics for each dataset, including accuracy, FPR, and recall. 
These results demonstrate that the proposed method effectively improves cyber-attack detection in smart grids, achieving high accuracy and low FPR across varying data distributions.
\begin{table}[!ht]
  \centering
  \caption{Performance of the Proposed Method on Datasets}
  \label{tab:tommoris}
  \begin{tabular}{ccccccc}
    \toprule
    Datasets & Precision & Recall & TPR & FPR & F1-score & Accuracy \\
    \midrule
    Dataset 1 & 0.98 & 0.99 & 0.99 & 0.02 & 0.99 & 0.99 \\
    Dataset 2 & 0.99 & 0.99 & 0.99 & 0.01 & 0.99 & 0.99 \\
    \bottomrule
  \end{tabular}
\end{table}
\subsection{Comparison with State of Art Methods}
We compare our proposed method with recent semi-supervised approaches in literature \cite{zhang2020detecting, dairi2023semi}.
Both these works generate synthetic labeled data to help the following anomaly detection module utilizing the \textit{generative adversarial networks}.
In \cite{zhang2020detecting}, the authors incorporated an autoencoder within an advanced GAN framework, which generates supervised data that aligns with the distribution of labeled data. 
Additionally, the autoencoder is trained to minimize supervised loss, enabling it to identify anomalies in the unlabeled data. 
In \cite{dairi2023semi}, a \textit{gated recurrent unit-based autoencoder} (AE-GRU) and \textit{gated recurrent unit-based GAN} (GAN-GRU) were combined with various anomaly detection techniques, including \textit{one-class SVM} (1SVM), \textit{Local Outlier Factor} (LOF), \textit{Isolation Forest} (iF), and \textit{Elliptic Envelope} (EE). 
Consequently, these methods are limited to binary classification, distinguishing only attack events. 
In contrast, the proposed method extends classification to unlabeled instances by identifying specific attack types that match those present in the labeled data. 
We also compare the proposed method with a label propagation-based approach \cite{zhang2021flexmatch}, which achieves a comparable accuracy of $98\%$ on the same group of datasets. However, the proposed method demonstrates superior efficiency in terms of computational time. 

The asymptotic training time for the approach in \cite{zhang2021flexmatch} is $(O(N))$, where $(N)$ represents the total number of labeled and unlabeled samples, while its prediction time for a new incoming sample is $(O(N_l))$, where $(N_l)$ is the number of labeled samples. 

In contrast, the proposed method has an asymptotic training time of $O(N_u)$, where $(N_u)$ is the number of unlabeled samples, and its prediction time for a new incoming sample is constant at $O(1)$. This significant difference in time complexity highlights the efficiency of the proposed method.
Table ~\ref{tab:perf} presents the comparison results with the state-of-the-art methods.
\begin{table}[!ht]
  \caption{Performance of the Methods}
  \label{tab:perf}
  \begin{tabular}{cccccl}
    \toprule
    Methods & Precision & Recall & FPR & F1-score & Accuracy \\
    \midrule
    Method in \cite{dairi2023semi} & 0.91 & 0.89 & 0.11 & 0.90 & 0.82 \\
    Method in \cite{zhang2020detecting} & 0.93 & 0.98 & 0.15 & 0.96 & 0.92 \\
    Method in \cite{zhang2021flexmatch} & 0.95 & 0.97 & 0.05 & 0.98 & 0.97 \\
    Proposed Method & 0.99 & 0.99 & 0.01 & 0.99 & 0.99 \\
  \bottomrule
\end{tabular}
\end{table}
\section{Conclusion}
The smart grids provide reliable power delivery by including advanced digital technologies. 
However, the adversaries exploit new security vulnerabilities to launch cyberattacks in to the smartgrids. These lead to the damage of potential power supply and management of the grid.  
Detecting these attacks is difficult when there is a scarcity of patterns of measurements during these scenarios. To address this concern in cyber-attack detection, we developed a semi-supervised scheme that uses a significantly small dataset of labeled patterns of measurements to classify large sets of unknown samples. While existing semi-supervised approaches successfully identify anomalies using GAN-generated data, they are typically restricted to binary classification, differentiating only between attack and normal events. The proposed method advances this by leveraging labeled data to classify unlabeled instances into distinct attack types, providing a more granular understanding of the anomalies present. This approach enhances anomaly detection and offers insight into specific attack categories, potentially improving response strategies in complex, multi-class anomaly environments with higher accuracy.
\bibliography{references}


\begin{thebibliography}{39}
\ifx \bisbn   \undefined \def \bisbn  #1{ISBN #1}\fi
\ifx \binits  \undefined \def \binits#1{#1}\fi
\ifx \bauthor  \undefined \def \bauthor#1{#1}\fi
\ifx \batitle  \undefined \def \batitle#1{#1}\fi
\ifx \bjtitle  \undefined \def \bjtitle#1{#1}\fi
\ifx \bvolume  \undefined \def \bvolume#1{\textbf{#1}}\fi
\ifx \byear  \undefined \def \byear#1{#1}\fi
\ifx \bissue  \undefined \def \bissue#1{#1}\fi
\ifx \bfpage  \undefined \def \bfpage#1{#1}\fi
\ifx \blpage  \undefined \def \blpage #1{#1}\fi
\ifx \burl  \undefined \def \burl#1{\textsf{#1}}\fi
\ifx \doiurl  \undefined \def \doiurl#1{\url{https://doi.org/#1}}\fi
\ifx \betal  \undefined \def \betal{\textit{et al.}}\fi
\ifx \binstitute  \undefined \def \binstitute#1{#1}\fi
\ifx \binstitutionaled  \undefined \def \binstitutionaled#1{#1}\fi
\ifx \bctitle  \undefined \def \bctitle#1{#1}\fi
\ifx \beditor  \undefined \def \beditor#1{#1}\fi
\ifx \bpublisher  \undefined \def \bpublisher#1{#1}\fi
\ifx \bbtitle  \undefined \def \bbtitle#1{#1}\fi
\ifx \bedition  \undefined \def \bedition#1{#1}\fi
\ifx \bseriesno  \undefined \def \bseriesno#1{#1}\fi
\ifx \blocation  \undefined \def \blocation#1{#1}\fi
\ifx \bsertitle  \undefined \def \bsertitle#1{#1}\fi
\ifx \bsnm \undefined \def \bsnm#1{#1}\fi
\ifx \bsuffix \undefined \def \bsuffix#1{#1}\fi
\ifx \bparticle \undefined \def \bparticle#1{#1}\fi
\ifx \barticle \undefined \def \barticle#1{#1}\fi
\bibcommenthead
\ifx \bconfdate \undefined \def \bconfdate #1{#1}\fi
\ifx \botherref \undefined \def \botherref #1{#1}\fi
\ifx \url \undefined \def \url#1{\textsf{#1}}\fi
\ifx \bchapter \undefined \def \bchapter#1{#1}\fi
\ifx \bbook \undefined \def \bbook#1{#1}\fi
\ifx \bcomment \undefined \def \bcomment#1{#1}\fi
\ifx \oauthor \undefined \def \oauthor#1{#1}\fi
\ifx \citeauthoryear \undefined \def \citeauthoryear#1{#1}\fi
\ifx \endbibitem  \undefined \def \endbibitem {}\fi
\ifx \bconflocation  \undefined \def \bconflocation#1{#1}\fi
\ifx \arxivurl  \undefined \def \arxivurl#1{\textsf{#1}}\fi
\csname PreBibitemsHook\endcsname

\bibitem[\protect\citeauthoryear{Mitchell and Chen}{2014}]{mitchell2014survey}
\begin{barticle}
\bauthor{\bsnm{Mitchell}, \binits{R.}},
\bauthor{\bsnm{Chen}, \binits{I.}}:
\batitle{A survey of intrusion detection techniques for cyber-physical systems}.
\bjtitle{ACM Computing Surveys (CSUR)}
\bvolume{46}(\bissue{4}),
\bfpage{1}--\blpage{29}
(\byear{2014})
\end{barticle}
\endbibitem

\bibitem[\protect\citeauthoryear{Yang et~al.}{2021}]{yang2021survey}
\begin{barticle}
\bauthor{\bsnm{Yang}, \binits{X.}},
\bauthor{\bsnm{Shu}, \binits{L.}},
\bauthor{\bsnm{Chen}, \binits{J.}},
\bauthor{\bsnm{Ferrag}, \binits{M.A.}},
\bauthor{\bsnm{Wu}, \binits{J.}},
\bauthor{\bsnm{Nurellari}, \binits{E.}},
\bauthor{\bsnm{Huang}, \binits{K.}}:
\batitle{A survey on smart agriculture: Development modes, technologies, and security and privacy challenges}.
\bjtitle{IEEE/CAA Journal of Automatica Sinica}
\bvolume{8}(\bissue{2}),
\bfpage{273}--\blpage{302}
(\byear{2021})
\end{barticle}
\endbibitem

\bibitem[\protect\citeauthoryear{Zhang et~al.}{2021}]{dlsurvey}
\begin{botherref}
\oauthor{\bsnm{Zhang}, \binits{J.}},
\oauthor{\bsnm{Pan}, \binits{L.}},
\oauthor{\bsnm{Han}, \binits{Q.L.}},
\oauthor{\bsnm{Chen}, \binits{C.}},
\oauthor{\bsnm{Wen}, \binits{S.}},
\oauthor{\bsnm{Xiang}, \binits{Y.}}:
Deep learning based attack detection for cyber-physical system cybersecurity: A survey.
IEEE/CAA Journal of Automatica Sinica
\textbf{9}(3)
(2021)
\end{botherref}
\endbibitem

\bibitem[\protect\citeauthoryear{Karnouskos}{2011}]{karnouskos2011stuxnet}
\begin{bchapter}
\bauthor{\bsnm{Karnouskos}, \binits{S.}}:
\bctitle{Stuxnet worm impact on industrial cyber-physical system security}.
In: \bbtitle{IECON 2011-37th Annual Conference of the IEEE Industrial Electronics Society},
pp. \bfpage{4490}--\blpage{4494}.
\bpublisher{IEEE},
\blocation{Melbourne, Victoria, Australia}
(\byear{2011})
\end{bchapter}
\endbibitem

\bibitem[\protect\citeauthoryear{Lehman and Maras}{2024}]{lehman15331cyber}
\begin{botherref}
\oauthor{\bsnm{Lehman}, \binits{G.}},
\oauthor{\bsnm{Maras}, \binits{P.}}:
Cyber-attack against ukrainian power plants.
Prykarpattyaoblenergo and Kyivoblenergo. Available online: https://nsarchive. gwu. edu/media/15331/ocr (accessed on 1 December 2024)
(2024)
\end{botherref}
\endbibitem

\bibitem[\protect\citeauthoryear{Liu et~al.}{2011}]{liu2011false}
\begin{barticle}
\bauthor{\bsnm{Liu}, \binits{Y.}},
\bauthor{\bsnm{Ning}, \binits{P.}},
\bauthor{\bsnm{Reiter}, \binits{M.K.}}:
\batitle{False data injection attacks against state estimation in electric power grids}.
\bjtitle{ACM Transactions on Information and System Security (TISSEC)}
\bvolume{14}(\bissue{1}),
\bfpage{1}--\blpage{33}
(\byear{2011})
\end{barticle}
\endbibitem

\bibitem[\protect\citeauthoryear{Kosut et~al.}{2011}]{kosut2011malicious}
\begin{barticle}
\bauthor{\bsnm{Kosut}, \binits{O.}},
\bauthor{\bsnm{Jia}, \binits{L.}},
\bauthor{\bsnm{Thomas}, \binits{R.J.}},
\bauthor{\bsnm{Tong}, \binits{L.}}:
\batitle{Malicious data attacks on the smart grid}.
\bjtitle{IEEE Transactions on Smart Grid}
\bvolume{2}(\bissue{4}),
\bfpage{645}--\blpage{658}
(\byear{2011})
\end{barticle}
\endbibitem

\bibitem[\protect\citeauthoryear{Rawat and Bajracharya}{2015}]{rawat2015detection}
\begin{barticle}
\bauthor{\bsnm{Rawat}, \binits{D.B.}},
\bauthor{\bsnm{Bajracharya}, \binits{C.}}:
\batitle{Detection of false data injection attacks in smart grid communication systems}.
\bjtitle{IEEE Signal Processing Letters}
\bvolume{22}(\bissue{10}),
\bfpage{1652}--\blpage{1656}
(\byear{2015})
\end{barticle}
\endbibitem

\bibitem[\protect\citeauthoryear{Khan et~al.}{2017}]{khan2017model}
\begin{bchapter}
\bauthor{\bsnm{Khan}, \binits{R.}},
\bauthor{\bsnm{Albalushi}, \binits{A.}},
\bauthor{\bsnm{McLaughlin}, \binits{K.}},
\bauthor{\bsnm{Laverty}, \binits{D.}},
\bauthor{\bsnm{Sezer}, \binits{S.}}:
\bctitle{Model based intrusion detection system for synchrophasor applications in smart grid}.
In: \bbtitle{2017 IEEE Power \& Energy Society General Meeting},
pp. \bfpage{1}--\blpage{5}.
\bpublisher{IEEE},
\blocation{Chicago, IL USA}
(\byear{2017})
\end{bchapter}
\endbibitem

\bibitem[\protect\citeauthoryear{Zhang et~al.}{2020}]{zhang2020detecting}
\begin{barticle}
\bauthor{\bsnm{Zhang}, \binits{Y.}},
\bauthor{\bsnm{Wang}, \binits{J.}},
\bauthor{\bsnm{Chen}, \binits{B.}}:
\batitle{Detecting false data injection attacks in smart grids: A semi-supervised deep learning approach}.
\bjtitle{IEEE Transactions on Smart Grid}
\bvolume{12}(\bissue{1}),
\bfpage{623}--\blpage{634}
(\byear{2020})
\end{barticle}
\endbibitem

\bibitem[\protect\citeauthoryear{Musleh et~al.}{2019}]{musleh2019survey}
\begin{barticle}
\bauthor{\bsnm{Musleh}, \binits{A.S.}},
\bauthor{\bsnm{Chen}, \binits{G.}},
\bauthor{\bsnm{Dong}, \binits{Z.Y.}}:
\batitle{A survey on the detection algorithms for false data injection attacks in smart grids}.
\bjtitle{IEEE Transactions on Smart Grid}
\bvolume{11}(\bissue{3}),
\bfpage{2218}--\blpage{2234}
(\byear{2019})
\end{barticle}
\endbibitem

\bibitem[\protect\citeauthoryear{Acosta et~al.}{2020}]{acosta2020extremely}
\begin{barticle}
\bauthor{\bsnm{Acosta}, \binits{M.R.C.}},
\bauthor{\bsnm{Ahmed}, \binits{S.}},
\bauthor{\bsnm{Garcia}, \binits{C.E.}},
\bauthor{\bsnm{Koo}, \binits{I.}}:
\batitle{Extremely randomized trees-based scheme for stealthy cyber-attack detection in smart grid networks}.
\bjtitle{IEEE access}
\bvolume{8},
\bfpage{19921}--\blpage{19933}
(\byear{2020})
\end{barticle}
\endbibitem

\bibitem[\protect\citeauthoryear{Sakhnini et~al.}{2019}]{sakhnini2019smart}
\begin{bchapter}
\bauthor{\bsnm{Sakhnini}, \binits{J.}},
\bauthor{\bsnm{Karimipour}, \binits{H.}},
\bauthor{\bsnm{Dehghantanha}, \binits{A.}}:
\bctitle{Smart grid cyber attacks detection using supervised learning and heuristic feature selection}.
In: \bbtitle{2019 IEEE 7th International Conference on Smart Energy Grid Engineering (SEGE)},
pp. \bfpage{108}--\blpage{112}.
\bpublisher{IEEE},
\blocation{Oshawa, Ontario, Canada}
(\byear{2019})
\end{bchapter}
\endbibitem

\bibitem[\protect\citeauthoryear{Niu et~al.}{2019}]{niu2019dynamic}
\begin{bchapter}
\bauthor{\bsnm{Niu}, \binits{X.}},
\bauthor{\bsnm{Li}, \binits{J.}},
\bauthor{\bsnm{Sun}, \binits{J.}},
\bauthor{\bsnm{Tomsovic}, \binits{K.}}:
\bctitle{Dynamic detection of false data injection attack in smart grid using deep learning}.
In: \bbtitle{2019 IEEE Power \& Energy Society Innovative Smart Grid Technologies Conference (ISGT)},
pp. \bfpage{1}--\blpage{6}.
\bpublisher{IEEE},
\blocation{Washington D.C.}
(\byear{2019})
\end{bchapter}
\endbibitem

\bibitem[\protect\citeauthoryear{Bitirgen and Filik}{2023}]{bitirgen2023hybrid}
\begin{barticle}
\bauthor{\bsnm{Bitirgen}, \binits{K.}},
\bauthor{\bsnm{Filik}, \binits{{\"U}.B.}}:
\batitle{A hybrid deep learning model for discrimination of physical disturbance and cyber-attack detection in smart grid}.
\bjtitle{International Journal of Critical Infrastructure Protection}
\bvolume{40},
\bfpage{100582}
(\byear{2023})
\end{barticle}
\endbibitem

\bibitem[\protect\citeauthoryear{Maglaras and Jiang}{2014}]{maglaras2014ocsvm}
\begin{bchapter}
\bauthor{\bsnm{Maglaras}, \binits{L.A.}},
\bauthor{\bsnm{Jiang}, \binits{J.}}:
\bctitle{Ocsvm model combined with k-means recursive clustering for intrusion detection in scada systems}.
In: \bbtitle{10th International Conference on Heterogeneous Networking for Quality, Reliability, Security and Robustness},
pp. \bfpage{133}--\blpage{134}.
\bpublisher{IEEE},
\blocation{Rhodes, Greece}
(\byear{2014})
\end{bchapter}
\endbibitem

\bibitem[\protect\citeauthoryear{Dairi et~al.}{2023}]{dairi2023semi}
\begin{bchapter}
\bauthor{\bsnm{Dairi}, \binits{A.}},
\bauthor{\bsnm{Harrou}, \binits{F.}},
\bauthor{\bsnm{Bouyeddou}, \binits{B.}},
\bauthor{\bsnm{Senouci}, \binits{S.M.}},
\bauthor{\bsnm{Sun}, \binits{Y.}}:
\bctitle{Semi-supervised deep learning-driven anomaly detection schemes for cyber-attack detection in smart grids}.
In: \bbtitle{Power Systems Cybersecurity: Methods, Concepts, and Best Practices},
pp. \bfpage{265}--\blpage{295}.
\bpublisher{Springer},
\blocation{Cham, Switzerland}
(\byear{2023})
\end{bchapter}
\endbibitem

\bibitem[\protect\citeauthoryear{Lee}{2013}]{lee2013pseudo}
\begin{bchapter}
\bauthor{\bsnm{Lee}, \binits{D.H.}}:
\bctitle{Pseudo-label: The simple and efficient semi-supervised learning method for deep neural networks}.
In: \bbtitle{Workshop on Challenges in Representation Learning, ICML},
vol. \bseriesno{3},
p. \bfpage{896}.
\bpublisher{JMLR: Workshop and Conference Proceedings},
\blocation{Atlanta, Georgia, USA}
(\byear{2013})
\end{bchapter}
\endbibitem

\bibitem[\protect\citeauthoryear{Fan et~al.}{2023}]{fan2023revisiting}
\begin{barticle}
\bauthor{\bsnm{Fan}, \binits{Y.}},
\bauthor{\bsnm{Kukleva}, \binits{A.}},
\bauthor{\bsnm{Dai}, \binits{D.}},
\bauthor{\bsnm{Schiele}, \binits{B.}}:
\batitle{Revisiting consistency regularization for semi-supervised learning}.
\bjtitle{International Journal of Computer Vision}
\bvolume{131}(\bissue{3}),
\bfpage{626}--\blpage{643}
(\byear{2023})
\end{barticle}
\endbibitem

\bibitem[\protect\citeauthoryear{Tang et~al.}{2023}]{tang2023towards}
\begin{barticle}
\bauthor{\bsnm{Tang}, \binits{Y.}},
\bauthor{\bsnm{Ge}, \binits{J.}},
\bauthor{\bsnm{Guo}, \binits{K.}},
\bauthor{\bsnm{Zheng}, \binits{Y.}},
\bauthor{\bsnm{Hu}, \binits{H.}},
\bauthor{\bsnm{Liang}, \binits{J.}}:
\batitle{Towards better utilization of pseudo labels for weakly supervised temporal action localization}.
\bjtitle{Information Sciences}
\bvolume{623},
\bfpage{693}--\blpage{708}
(\byear{2023})
\end{barticle}
\endbibitem

\bibitem[\protect\citeauthoryear{Bengio et~al.}{2009}]{bengio2009curriculum}
\begin{bchapter}
\bauthor{\bsnm{Bengio}, \binits{Y.}},
\bauthor{\bsnm{Louradour}, \binits{J.}},
\bauthor{\bsnm{Collobert}, \binits{R.}},
\bauthor{\bsnm{Weston}, \binits{J.}}:
\bctitle{Curriculum learning}.
In: \bbtitle{Proceedings of the 26th Annual International Conference on Machine Learning},
pp. \bfpage{41}--\blpage{48}.
\bpublisher{ACM},
\blocation{Montreal, Canada}
(\byear{2009})
\end{bchapter}
\endbibitem

\bibitem[\protect\citeauthoryear{Liu et~al.}{2017}]{liu2017review}
\begin{barticle}
\bauthor{\bsnm{Liu}, \binits{Y.}},
\bauthor{\bsnm{Peng}, \binits{Y.}},
\bauthor{\bsnm{Wang}, \binits{B.}},
\bauthor{\bsnm{Yao}, \binits{S.}},
\bauthor{\bsnm{Liu}, \binits{Z.}}:
\batitle{Review on cyber-physical systems}.
\bjtitle{IEEE/CAA Journal of Automatica Sinica}
\bvolume{4}(\bissue{1}),
\bfpage{27}--\blpage{40}
(\byear{2017})
\end{barticle}
\endbibitem

\bibitem[\protect\citeauthoryear{Gunduz and Das}{2018}]{gunduz2018analysis}
\begin{bchapter}
\bauthor{\bsnm{Gunduz}, \binits{M.Z.}},
\bauthor{\bsnm{Das}, \binits{R.}}:
\bctitle{Analysis of cyber-attacks on smart grid applications}.
In: \bbtitle{2018 International Conference on Artificial Intelligence and Data Processing (IDAP)},
pp. \bfpage{1}--\blpage{5}.
\bpublisher{IEEE},
\blocation{Malatya, Turkey}
(\byear{2018})
\end{bchapter}
\endbibitem

\bibitem[\protect\citeauthoryear{Sohn et~al.}{2020}]{sohn2020fixmatch}
\begin{barticle}
\bauthor{\bsnm{Sohn}, \binits{K.}},
\bauthor{\bsnm{Berthelot}, \binits{D.}},
\bauthor{\bsnm{Carlini}, \binits{N.}},
\bauthor{\bsnm{Zhang}, \binits{Z.}},
\bauthor{\bsnm{Zhang}, \binits{H.}},
\bauthor{\bsnm{Raffel}, \binits{C.A.}},
\bauthor{\bsnm{Cubuk}, \binits{E.D.}},
\bauthor{\bsnm{Kurakin}, \binits{A.}},
\bauthor{\bsnm{Li}, \binits{C.L.}}:
\batitle{Fixmatch: Simplifying semi-supervised learning with consistency and confidence}.
\bjtitle{Advances in neural information processing systems}
\bvolume{33},
\bfpage{596}--\blpage{608}
(\byear{2020})
\end{barticle}
\endbibitem

\bibitem[\protect\citeauthoryear{Arazo et~al.}{2020}]{arazo2020pseudo}
\begin{bchapter}
\bauthor{\bsnm{Arazo}, \binits{E.}},
\bauthor{\bsnm{Ortego}, \binits{D.}},
\bauthor{\bsnm{Albert}, \binits{P.}},
\bauthor{\bsnm{O’Connor}, \binits{N.E.}},
\bauthor{\bsnm{McGuinness}, \binits{K.}}:
\bctitle{Pseudo-labeling and confirmation bias in deep semi-supervised learning}.
In: \bbtitle{2020 International Joint Conference on Neural Networks (IJCNN)},
pp. \bfpage{1}--\blpage{8}.
\bpublisher{IEEE},
\blocation{Glasgow, United Kingdom}
(\byear{2020})
\end{bchapter}
\endbibitem

\bibitem[\protect\citeauthoryear{Zhang et~al.}{2021}]{zhang2021flexmatch}
\begin{barticle}
\bauthor{\bsnm{Zhang}, \binits{B.}},
\bauthor{\bsnm{Wang}, \binits{Y.}},
\bauthor{\bsnm{Hou}, \binits{W.}},
\bauthor{\bsnm{Wu}, \binits{H.}},
\bauthor{\bsnm{Wang}, \binits{J.}},
\bauthor{\bsnm{Okumura}, \binits{M.}},
\bauthor{\bsnm{Shinozaki}, \binits{T.}}:
\batitle{Flexmatch: Boosting semi-supervised learning with curriculum pseudo labeling}.
\bjtitle{Advances in Neural Information Processing Systems}
\bvolume{34},
\bfpage{18408}--\blpage{18419}
(\byear{2021})
\end{barticle}
\endbibitem

\bibitem[\protect\citeauthoryear{Li et~al.}{2019}]{li2019naive}
\begin{barticle}
\bauthor{\bsnm{Li}, \binits{Z.}},
\bauthor{\bsnm{Ko}, \binits{B.}},
\bauthor{\bsnm{Choi}, \binits{H.}}:
\batitle{Naive semi-supervised deep learning using pseudo-label}.
\bjtitle{Peer-to-peer networking and applications}
\bvolume{12},
\bfpage{1358}--\blpage{1368}
(\byear{2019})
\end{barticle}
\endbibitem

\bibitem[\protect\citeauthoryear{Shi et~al.}{2018}]{shi2018transductive}
\begin{bchapter}
\bauthor{\bsnm{Shi}, \binits{W.}},
\bauthor{\bsnm{Gong}, \binits{Y.}},
\bauthor{\bsnm{Ding}, \binits{C.}},
\bauthor{\bsnm{Tao}, \binits{Z.M.}},
\bauthor{\bsnm{Zheng}, \binits{N.}}:
\bctitle{Transductive semi-supervised deep learning using min-max features}.
In: \bbtitle{Proceedings of the European Conference on Computer Vision (ECCV)},
pp. \bfpage{299}--\blpage{315}
(\byear{2018})
\end{bchapter}
\endbibitem

\bibitem[\protect\citeauthoryear{Pelletier et~al.}{2019}]{pelletier2019temporal}
\begin{barticle}
\bauthor{\bsnm{Pelletier}, \binits{C.}},
\bauthor{\bsnm{Webb}, \binits{G.I.}},
\bauthor{\bsnm{Petitjean}, \binits{F.}}:
\batitle{Temporal convolutional neural network for the classification of satellite image time series}.
\bjtitle{Remote Sensing}
\bvolume{11}(\bissue{5}),
\bfpage{523}
(\byear{2019})
\end{barticle}
\endbibitem

\bibitem[\protect\citeauthoryear{Lin et~al.}{2020}]{lin2020temporal}
\begin{bchapter}
\bauthor{\bsnm{Lin}, \binits{Y.}},
\bauthor{\bsnm{Koprinska}, \binits{I.}},
\bauthor{\bsnm{Rana}, \binits{M.}}:
\bctitle{Temporal convolutional neural networks for solar power forecasting}.
In: \bbtitle{2020 International Joint Conference on Neural Networks (IJCNN)},
pp. \bfpage{1}--\blpage{8}.
\bpublisher{IEEE},
\blocation{Glasgow, United Kingdom}
(\byear{2020})
\end{bchapter}
\endbibitem

\bibitem[\protect\citeauthoryear{Kiranyaz et~al.}{2021}]{kiranyaz20211d}
\begin{barticle}
\bauthor{\bsnm{Kiranyaz}, \binits{S.}},
\bauthor{\bsnm{Avci}, \binits{O.}},
\bauthor{\bsnm{Abdeljaber}, \binits{O.}},
\bauthor{\bsnm{Ince}, \binits{T.}},
\bauthor{\bsnm{Gabbouj}, \binits{M.}},
\bauthor{\bsnm{Inman}, \binits{D.J.}}:
\batitle{1d convolutional neural networks and applications: A survey}.
\bjtitle{Mechanical systems and signal processing}
\bvolume{151},
\bfpage{107398}
(\byear{2021})
\end{barticle}
\endbibitem

\bibitem[\protect\citeauthoryear{Matoušek et~al.}{2022}]{1trw-n685-22}
\begin{botherref}
\oauthor{\bsnm{Matoušek}, \binits{P.}},
\oauthor{\bsnm{Ryšavý}, \binits{O.}},
\oauthor{\bsnm{Grofčík}, \binits{P.}}:
ICS Dataset for Smart Grid Anomaly Detection.
\doiurl{10.21227/1trw-n685} .
\url{https://dx.doi.org/10.21227/1trw-n685}
\end{botherref}
\endbibitem

\bibitem[\protect\citeauthoryear{Morris and Gao}{2014}]{morris2014industrial}
\begin{bchapter}
\bauthor{\bsnm{Morris}, \binits{T.H.}},
\bauthor{\bsnm{Gao}, \binits{W.}}:
\bctitle{Industrial control system traffic data sets for intrusion detection research}.
In: \bbtitle{Critical Infrastructure Protection VIII: 8th IFIP WG 11.10 International Conference, ICCIP},
pp. \bfpage{65}--\blpage{78}.
\bpublisher{Springer},
\blocation{Arlington, VA, USA}
(\byear{2014})
\end{bchapter}
\endbibitem

\bibitem[\protect\citeauthoryear{Matou{\v{s}}ek et~al.}{2020}]{matouvsek2020flow}
\begin{barticle}
\bauthor{\bsnm{Matou{\v{s}}ek}, \binits{P.}},
\bauthor{\bsnm{Ry{\v{s}}av{\`y}}, \binits{O.}},
\bauthor{\bsnm{Gr{\'e}gr}, \binits{M.}},
\bauthor{\bsnm{Havlena}, \binits{V.}}:
\batitle{Flow-based monitoring of ics communication in the smart grid}.
\bjtitle{Journal of Information Security and Applications}
\bvolume{54},
\bfpage{102535}
(\byear{2020})
\end{barticle}
\endbibitem

\bibitem[\protect\citeauthoryear{Matou{\v{s}}ek et~al.}{2021}]{matouvsek2021efficient}
\begin{bchapter}
\bauthor{\bsnm{Matou{\v{s}}ek}, \binits{P.}},
\bauthor{\bsnm{Havlena}, \binits{V.}},
\bauthor{\bsnm{Hol{\'\i}k}, \binits{L.}}:
\bctitle{Efficient modelling of ics communication for anomaly detection using probabilistic automata}.
In: \bbtitle{2021 IFIP/IEEE International Symposium on Integrated Network Management (IM)},
pp. \bfpage{81}--\blpage{89}.
\bpublisher{IEEE},
\blocation{Bordeaux, France}
(\byear{2021})
\end{bchapter}
\endbibitem

\bibitem[\protect\citeauthoryear{T.H. et~al.}{2014}]{tommymorris}
\begin{botherref}
\oauthor{\bsnm{T.H.}, \binits{M.}},
\oauthor{\bsnm{U.}, \binits{A.}},
\oauthor{\bsnm{S.}, \binits{P.}},
\oauthor{\bsnm{R.}, \binits{B.}},
\oauthor{\bsnm{J.}, \binits{B.}}:
ICS Dataset for Smart Grid Anomaly Detection.
\url{https://sites.google.com/a/uah.edu/tommy-morris-uah/ics-data-sets?authuser=0}
\end{botherref}
\endbibitem

\bibitem[\protect\citeauthoryear{Hink et~al.}{2014}]{hink2014machine}
\begin{bchapter}
\bauthor{\bsnm{Hink}, \binits{R.C.B.}},
\bauthor{\bsnm{Beaver}, \binits{J.M.}},
\bauthor{\bsnm{Buckner}, \binits{M.A.}},
\bauthor{\bsnm{Morris}, \binits{T.}},
\bauthor{\bsnm{Adhikari}, \binits{U.}},
\bauthor{\bsnm{Pan}, \binits{S.}}:
\bctitle{Machine learning for power system disturbance and cyber-attack discrimination}.
In: \bbtitle{2014 7th International Symposium on Resilient Control Systems (ISRCS)},
pp. \bfpage{1}--\blpage{8}.
\bpublisher{IEEE},
\blocation{Denver, Colorado}
(\byear{2014})
\end{bchapter}
\endbibitem

\bibitem[\protect\citeauthoryear{Hossin and Sulaiman}{2015}]{hossin2015review}
\begin{barticle}
\bauthor{\bsnm{Hossin}, \binits{M.}},
\bauthor{\bsnm{Sulaiman}, \binits{M.N.}}:
\batitle{A review on evaluation metrics for data classification evaluations}.
\bjtitle{International journal of data mining \& knowledge management process}
\bvolume{5}(\bissue{2}),
\bfpage{1}
(\byear{2015})
\end{barticle}
\endbibitem

\bibitem[\protect\citeauthoryear{Tharwat}{2021}]{tharwat2021classification}
\begin{barticle}
\bauthor{\bsnm{Tharwat}, \binits{A.}}:
\batitle{Classification assessment methods}.
\bjtitle{Applied computing and informatics}
\bvolume{17}(\bissue{1}),
\bfpage{168}--\blpage{192}
(\byear{2021})
\end{barticle}
\endbibitem

\end{thebibliography}
\end{document}